\documentstyle[12pt,epsf]{article}

\newcommand{\be}{\begin{equation}}
\newcommand{\ee}{\end{equation}}
\newcommand{\bea}{\begin{eqnarray}}
\newcommand{\eea}{\end{eqnarray}}
\newcommand{\ena}{\end{eqnarray}}
\newcommand{\vs}[1]{\vspace{#1 mm}}
\newcommand{\hs}[1]{\hspace{#1 mm}}
\newcommand{\ba}{\begin{array}}
\newcommand{\ea}{\end{array}}
\newcommand{\beann}{\begin{eqnarray*}}
\newcommand{\eeann}{\end{eqnarray*}}
\newcommand{\nn}{\nonumber \\}
\renewcommand{\a}{\alpha}
\renewcommand{\b}{\beta}
\renewcommand{\c}{\gamma}
\renewcommand{\d}{\delta}
\newcommand{\e}{\epsilon}

\newcommand{\diag}{{\rm diag}}

\newcommand{\del}{\partial}
\newcommand{\pa}{\partial}
\newcommand{\one}{{\mathbf 1}}
\newcommand{\Z}{{\mathbf Z}}
\newcommand{\R}{{\mathbf R}}
\newcommand{\mb}[1]{\mbox{\boldmath $#1$}}
\newcommand{\rot}[3]{\left[{#1}\atop{#2}\right]_{#3}}

\makeatletter
  
  \@addtoreset{equation}{section}
\makeatother

\begin{document}
\topmargin 0pt
\oddsidemargin 0mm
\renewcommand{\thefootnote}{\fnsymbol{footnote}}

%
%
\begin{titlepage}

\setcounter{page}{0}
\begin{flushright}
UT-Komaba/98-19 \\
YITP-98-55 \\
OU-HET 302 \\
hep-th/9808111
\end{flushright}

\vs{7}
\begin{center}
{\Large\bf
Three-Dimensional Gauge Dynamics from Brane Configurations
with $(p,q)$-Fivebrane
}

\vs{15}
{\large
Takuhiro Kitao,$^{a,}$\footnote{e-mail address: kitao@hep1.c.u-tokyo.ac.jp}
Kazutoshi Ohta$^{b,}$\footnote{e-mail address:
kohta@yukawa.kyoto-u.ac.jp. Yukawa memorial fellow.}
and
Nobuyoshi Ohta$^{c,}$\footnote{e-mail address:
ohta@phys.wani.osaka-u.ac.jp}} \\
\vs{8}
$^a${\em Institute of Physics, University of Tokyo,
Komaba, Tokyo 153-8902, Japan} \\
\vs{2}
$^b${\em Yukawa Institute for Theoretical Physics, Kyoto University,
Kyoto 606-8502, Japan} \\
\vs{2}
$^c${\em Department of Physics, Osaka University,
Toyonaka, Osaka 560-0043, Japan} \\
\end{center}
\vs{10}
\centerline{{\bf{Abstract}}}
\vs{5}

We study three-dimensional gauge dynamics by using type IIB superstring brane
configurations, which can be obtained from the M-theory configuration of
M2-branes stretched between two M5-branes with relative angles.
Our construction of brane configurations includes $(p,q)$5-brane and gives
a systematic classification of possible three-dimensional gauge theories.
The explicit identification of gauge theories are made and their mirror
symmetry is discussed. As a new feature, our theories include interesting
Maxwell-Chern-Simons system whose vacuum structure is also examined in detail,
obtaining results consistent with the brane picture.

\end{titlepage}

\newpage
\renewcommand{\thefootnote}{\arabic{footnote}}
\setcounter{footnote}{0}

\section{Introduction}

Understanding nonperturbative properties of field theories is one of the
long-standing problems in particle physics. Recently there has been
much progress in this area. Many interesting and exact results on
supersymmetric gauge field theories in 3 (and other) dimensions have been
obtained in refs.~\cite{SW,IS,BHO1}. These results were then interpreted
and understood by analyzing the effective theory on the common worldvolume
of branes in type IIB superstring theory, in which D3-branes are suspended
between two parallel NS5-branes~\cite{PZ,HW,BHO2,BHO3,EGKRS}.
The ``mirror symmetry'' between the Coulomb and Higgs branches which
is miraculous from the field theory point of view was simply understood
as a known $SL(2,\Z)$ duality symmetry in the type IIB string
picture.\footnote{For a review and more references, see \cite{GK}.}

These results were soon generalized to four-dimensional theories by using
brane configurations in Type IIA superstrings~\cite{EGK,BA,W1}. Here again
the remarkable nonabelian duality in these theories are understood in terms
of brane exchange and creation phenomena first noted in ref.~\cite{HW}.
The four-dimensional $N{=}2$ theories also allow interpretation in terms
of the M-theory 5-brane solutions (M5-brane)~\cite{W1}. In fact other
brane configurations can also be understood as theories realized on the
rotated BPS brane configurations. Upon compactification and T-dualities
of M-brane configurations, many of these results can be described in
a unified manner.

In Type IIB theory, there also exists the so-called $(p,q)$5-brane which is
a bound state of $p$ D5-branes and $q$ NS5-branes. Due to its complicated
properties, this interesting solution has not been exploited in the above
context in the literature. However, from the M-theory point of view this
brane is simple and is nothing but an M5-brane wrapping the 11-th
direction~\cite{SCH}. This immediately leads to the possibility of
analyzing field theories realized on brane configurations with
$(p,q)$5-brane. The purpose of this paper is to classify possible theories
on the branes and discuss their properties when one of the NS5-branes
in Hanany-Witten configuration is rotated to the $(p,q)$5-brane. This
corresponds to the configuration of two M5-branes one of which is rotated
in the 11-th direction.

The question is then what kind of BPS solutions are possible. From the
11-dimensional point of view, this is reduced to the task of classifying
possible BPS configurations of two tilted M5-branes. Fortunately
this problem has been recently examined and the possible solutions are
sorted out in ref.~\cite{OT}. We will use this result to exhaust all
possible gauge theories for various rotations. To proceed, we introduce
M-branes between the two M5-branes. In ref.~\cite{W1}, the same M5-branes
were considered for such objects. In order to incorporate various rotation
angles in our case, however, we can have only the M2-branes. We then find
that possible theories are two- or three-dimensional gauge theories in
type IIA or IIB string theories, respectively, with various supersymmetries.

In this article we will concentrate on three-dimensional gauge theories in
type IIB superstrings. It turns out that there are various interesting theories
with $N=1,2,3,4$ supersymmetry. In particular, we find that the presence of
$(p,q)$5-brane requires the Chern-Simons (CS) terms in the gauge
theories.\footnote{Supersymmetric CS terms were first discussed in
ref.~\cite{Si} and later generalized in refs.~\cite{ZP,KL1,Iv}.}
We give an exhaustive list of possible theories and discuss the mirror
symmetry. The brane picture suggests that the vacuum structure of the
Maxwell-Chern-Simons (MCS) theories is quite rich, and we confirm this
from the field theory point of view.

The realization of odd numbers of supersymmetry, especially $N{=}3$, is also
an interesting feature of our new brane construction.\footnote{Supergravity
solution with this supersymmetry and the possible presence of the CS terms
in the effective field theories were discussed in ref.~\cite{GGPT}.
Realization of the solution in Matrix Theory is given in ref.~\cite{OZ}.}
It seems that very little is known about such theories and we hope that
our discussion sheds some light on this subject and stimulates further
investigations.

The CS theories has attracted much attention for their possible applications
to statistical physics, such as fractional quantum Hall effect and
superconductivity~\cite{IL,PO,CKS}. Our brane configurations give
a first realization of these theories in terms of branes and should be
useful to study exact properties of these theories.

The organization of the paper is as follows.
In sect.~2, we identify the BPS brane configurations in M- and string theories.
In sect.~2.1 we classify the BPS configurations consisting of relatively
rotated two M5-branes with $N_c$ M2-branes in between as well as $N_f$
M5-branes in 11-dimensional supergravity.
These are then related to brane configurations in type IIB
(and IIA) theory in sect.~2.2. In sect.~3, several arguments are given why
the gauge field theories contain CS terms for the configuration with
$(p,q)$5-brane. With these results, explicit identification of gauge
theories realized on the branes is made in sect.~4, where mirror theories
are also discussed. Further discussions of more nontrivial mirror symmetry
and the vacuum structure of the CS theories are given in sect.~5. In particular,
the quantum ground states of the MCS theory with rational coefficient for
the CS terms is discussed in detail by using its duality to self-dual models,
giving results consistent with the brane picture.
Sect.~6 is devoted to summary of
our results and discussions of remaining and future problems.

\section{Brane configurations with $(p,q)$5-brane}

The residual supersymmetry in the two M5-branes at angles has been examined
and classified in ref.~\cite{OT}. A natural question arises if such
configurations have any implications to field theories on the brane worldvolume.
Following ref.~\cite{HW}, we consider the configurations in which additional
branes exist between the two M5-branes. For such additional branes,
M5-branes were considered in ref.~\cite{W1}.
However, in order to allow for rotations with more than two angles, which are
of our interest, it turns out that M2-brane is the only possibility.
In this section, we first extend the results in ref.~\cite{OT} to include
the in-between M2-branes and find the conditions that there remains some
supersymmetry. These branes are then related to those in type IIB
superstrings.

\subsection{Brane configurations at angles in M-theory}

As in ref.~\cite{OT}, we start with two parallel M5-branes lying as
\bea
\begin{array}{lcccccccccc}
{\rm M}: & 1 & 2 & 3 & 4 & 5 & - & - & - & - & - \\
{\rm M}: & 1 & 2 & 3 & 4 & 5 & - & - & - & - & - 
\end{array}\, ,
\label{m5}
\ena
and consider fixing the first M5-brane but rotating the second one.
In order to have at least 2 dimensions in the intersection, we can only
rotate the second M5 by 4 angles. Suppose that these M5-branes are separated
in the direction of $x^6$. We find that we can include several M2-branes
between these M5-branes. Thus the worldvolumes for our brane configurations
in M-theory are summarized as
\bea
{\rm M5} &:& (12345), \nn
{\rm M2} &:& (1|6|), \nn
{\rm M5}'&:& \left(1\rot{2}{\natural}{\theta}\rot{3}{7}{\psi}
\rot{4}{8}{\varphi}\rot{5}{9}{\rho}\right),
\label{m52}
\ena
where the numbers denote the worldvolume directions (trivial 0 is not
exposed explicitly) and the vertical lines in M2-brane worldvolume implies
that the $x^6$ direction is terminated on the two M5-branes.
The vertical arrays in the second M5-brane worldvolume mean that the
directions are rotated by the angles indicated by subscripts.
Here we use the symbol $\natural$ for the number 10.

The presence of these branes imposes the constraints on the Killing
spinors~\cite{OT}
\bea
\label{c1}
{\rm M5} &:& \Gamma_{012345}\e = \e, \\
\label{c2}
{\rm M2} &:& \Gamma_{016}\e = \e,\\
\label{c3}
{\rm M5}'&:& R\Gamma_{012345}R^{-1}\e = \e,
\ena
where the rotation matrix for the second M5-brane is parametrized by
four angles as
\be
R=\exp\left\{
\frac{\theta}{2}\Gamma_{2\natural} + \frac{\psi}{2}\Gamma_{37}
+ \frac{\varphi}{2}\Gamma_{48}
+ \frac{\rho}{2}\Gamma_{59} \right\}.
\ee

Our task is to examine the simultaneous solutions of
eqs.~(\ref{c1})-(\ref{c3}) as functions of the four angles $\theta,
\psi, \varphi$ and $\rho$. Since $R\Gamma_{012345}R^{-1}=R^2\Gamma_{012345}$,
the third condition (\ref{c3}) is reduced to
\bea
(R^2-1)\e = 0.
\label{c3'}
\eea
A straightforward calculation gives
\bea
R^2-1 &=& 2R\Gamma_{2\natural}\left\{
\sin\frac{\theta}{2}\cos\frac{\psi}{2}\cos\frac{\varphi}{2}\cos\frac{\rho}{2}
-\Gamma_{2\natural 37}
\cos\frac{\theta}{2}\sin\frac{\psi}{2}\cos\frac{\varphi}{2}\cos\frac{\rho}{2}
\right. \nn
&&-\Gamma_{2\natural 48}
\cos\frac{\theta}{2}\cos\frac{\psi}{2}\sin\frac{\varphi}{2}\cos\frac{\rho}{2}
-\Gamma_{2\natural 59}
\cos\frac{\theta}{2}\cos\frac{\psi}{2}\cos\frac{\varphi}{2}\sin\frac{\rho}{2}
\nn
&&+\Gamma_{3748}
\sin\frac{\theta}{2}\sin\frac{\psi}{2}\sin\frac{\varphi}{2}\cos\frac{\rho}{2}
+\Gamma_{4859}
\sin\frac{\theta}{2}\cos\frac{\psi}{2}\sin\frac{\varphi}{2}\sin\frac{\rho}{2}
\nn
&& \left.+\Gamma_{3759}
\sin\frac{\theta}{2}\sin\frac{\psi}{2}\cos\frac{\varphi}{2}\sin\frac{\rho}{2}
-\Gamma_{2\natural 374859}
\cos\frac{\theta}{2}\sin\frac{\psi}{2}\sin\frac{\varphi}{2}\sin\frac{\rho}{2}
\right\}.
\eea

In order to find the solutions, it is convenient to choose
maximally diagonalized basis for the gamma matrices appearing in
these conditions. Since these products of gamma matrices commute with each
other and square to unity, and the traces of their products vanish,
we can arrange these matrices as
\bea
\Gamma_{012345} &=& \diag\left(\one_{16},-\one_{16}\right), \nn
\Gamma_{2\natural 37}&=&\diag\left(\one_{8},-\one_{8}, \cdots \right), \nn
\Gamma_{2\natural 48}&=&\diag\left(\one_{4},-\one_{4},\one_{4},-\one_{4},
 \cdots \right), \nn
\Gamma_{2\natural 59}&=&\diag\left(\one_{2},-\one_{2},\one_{2},
-\one_{2},\one_{2},-\one_{2},\one_{2},-\one_{2}, \cdots \right),
\label{ga}
\ena
where the rests are determined by those given and $\one_{n}$ stands for
$n \times n$ identity matrix.
In this basis, the first condition (\ref{c1}) kills the second 16 components
of the Killing spinor, and we have to examine the consequence of the second
(\ref{c2}) and third (\ref{c3'}) conditions for the first 16 components.

Using (\ref{ga}), we have
\bea
R^2-1 &=& 2R\Gamma_{2\natural}\nn
&&
\times\diag\left(
\sin\left(\frac{\theta-\psi-\varphi-\rho}{2}\right)\one_{2},
\sin\left(\frac{\theta-\psi-\varphi+\rho}{2}\right)\one_{2},
\right. \nn
&&
\sin\left(\frac{\theta-\psi+\varphi-\rho}{2}\right)\one_{2},
\sin\left(\frac{\theta-\psi+\varphi+\rho}{2}\right)\one_{2},\nn
&&
\sin\left(\frac{\theta+\psi-\varphi-\rho}{2}\right)\one_{2},
\sin\left(\frac{\theta+\psi-\varphi+\rho}{2}\right)\one_{2},\nn
&&
\left.
\sin\left(\frac{\theta+\psi+\varphi-\rho}{2}\right)\one_{2},
\sin\left(\frac{\theta+\psi+\varphi+\rho}{2}\right)\one_{2}, \cdots
\right),\nn
\Gamma_{016}-1 &=&
 -2\times\diag({\bf 0}_2,\one_{2},\one_{2},{\bf 0}_2,\one_{2},{\bf 0}_2,
{\bf 0}_2,\one_{2},\cdots).
\ena

Examining the various cases, we find the numbers of remaining supersymetry.
The results are summarized in table~\ref{t1} as the fraction of the original
32 supersymmetry of the M-theory vacuum. Also indicated in the table are the
numbers of supersymmetry counted as $d=3$ theories and the resulting branes
in type IIB superstring. These will be further discussed later on.
{\small
\begin{table}[htb]
\begin{tabular}{|c|c|c|c|c|l|}
\hline
& angles & condition & SUSY & d=3 & M5$'$ \\
\hline
\hline
1 &$\theta(2\natural)$ & $\theta=0$ & $\frac{1}{4}$ & $N{=}4$
 & NS5$\left(12345\right)$ \\
\hline
2-(i) & $\varphi(48), \rho(59)$ & $\rho=\varphi$ & $\frac{1}{8}$ & $N{=}2$
 & NS5$\left(123\rot{4}{8}{\varphi}\rot{5}{9}{\varphi}\right)$ \\
\hline
2-(ii) & $\theta(2\natural),\rho(59)$ & $\rho=\theta$ & $\frac{1}{8}$ & $N{=}2$
 & $(p,q)5\left(1234\rot{5}{9}{\theta}\right)$ \\
\hline
3-(i) & $\psi(37), \varphi(48), \rho(59)$ & $\rho=\psi+\varphi$ & $\frac{1}{16}$
 & $N{=}1$ & NS5$\left(12\rot{3}{7}{\psi}\rot{4}{8}{\varphi}
 \rot{5}{9}{\psi+\varphi}\right)$ \\
\hline
3-(ii)&  $\theta(2\natural), \varphi(48), \rho(59)$ & $\rho=\theta+\varphi$
 & $\frac{1}{16}$ & $N{=}1$ & $(p,q)5\left(123\rot{4}{8}{\varphi}
 \rot{5}{9}{\theta+\varphi}\right)$ \\
\hline
4-(i)& & $\rho=\theta+\psi+\varphi$ & $\frac{1}{16}$ & $N{=}1$ & $(p,q)5
\left(12\rot{3}{7}{\psi}\rot{4}{8}{\varphi}
\rot{5}{9}{\theta+\psi+\varphi}\right)$ \\
\cline{1-1}\cline{3-6}
4-(ii) & {\small $ \theta(2\natural), \psi(37), \varphi(48), \rho(59)$}
 & $\varphi=\psi, \rho=\theta$ & $\frac{1}{8}$ & $N{=}2$
 & $(p,q)5\left(12\rot{3}{7}{\psi}\rot{4}{8}{\psi}\rot{5}{9}{\theta}\right)$
 \\
\cline{1-1}\cline{3-6}
4-(iii) & & $\theta=\psi=\varphi=-\rho$ & $\frac{3}{16}$ & $N{=}3$
 & $(p,q)5\left(12\rot{3}{7}{\theta}\rot{4}{8}{\theta}
 \rot{5}{9}{-\theta}\right)$ \\
\hline
\end{tabular}
\caption{\small
Brane configurations at angles and various supersymmetric
theories in 3 dimensions.
}
\label{t1}
\end{table}
}

It is interesting to note that the remaining supersymmetry is reduced by
$\frac{1}{2}$ compared with the configuration without the intermediate
M2-branes. For example, the two-angle case in the second row in table 1
keeps $\frac{1}{4}$ supersymmetry for two M5-branes~\cite{OT}, and the
number is reduced by $\frac{1}{2}$ owing to the presence of the additional
M2-branes. However, the four-angle case is exceptional. In this case,
the previous analysis without M2-brane~\cite{OT} indicates that the
supersymmetry is the same as given in table~\ref{t1}, and so the number
of supersymmetry is not reduced only for this case.

Finally we note that the configurations considered above further allow
additional M5-branes lying in the directions
\bea
\begin{array}{lcccccccccc}
{\rm M}: & 1 & - & - & - & - & - & 7 & 8 & 9 & \natural \\
\end{array}\, .
\label{mat}
\ena
In fact the presence of this brane gives the constraint on the Killing spinor:
\bea
{\rm M5} &:& \Gamma_{01789\natural}\e = \e.
\label{c4}
\ena
Since the product of all gamma matrices is unity,
$\Gamma_{0123456789\natural}=1$, we have $\Gamma_{01789\natural}=
\Gamma_{016} \Gamma_{012345}$. Thus the constraint~(\ref{c4}) is reduced
to the combination of the conditions (\ref{c1}) and (\ref{c2}) without
producing any further restriction.
These additional M5-branes can serve as matter fields for the field theories.
We will consider the configurations in which the numbers of M2- and matter
M5-branes are $N_c$ and $N_f$, respectively.

\subsection{Brane configurations at angles in IIA and IIB string theories}

We now discuss compactification and T-duality of the previous brane
configurations, producing configurations in type IIA and IIB string theories.

Let us first consider compactifying in the direction of 11-th dimension,
and then making T-duality in the direction of $x^2$. We
suppose that the coordinates $(x^0,x^1,x^2,x^{10})$ has the topology of
$\R^{2}\times T^2$, the radii of the latter being $R_2$ and $R_{10}$,
respectively.

Under the compactification, M5-branes reduce to either NS5-brane or D4-brane
depending on their worldvolumes, and further T-dual transformation gives
the following branes:
\bea
\begin{tabular}{cccc}
{\rm M}: & M5(12345) & M5($1\natural 345$) & M5
$\left(1345\rot{2}{\natural}{\theta}\right)$ \\
& $\Downarrow$ & $\Downarrow$ & $\Downarrow$ \\
{\rm IIA}: & NS5(12345) & D4(1345) & NS5 $\oplus$ D4 \\
$T_2$-dual & $\Downarrow$ & $\Downarrow$ & $\Downarrow$ \\
{\rm IIB}: & NS5(12345) & D5(12345) & (p,q)5(12345)
\end{tabular}.
\label{tdual}
\ena
We note that the correspondence we are considering is schematically
written as $M/T^2 \simeq {\rm IIB}/S^1$.

\begin{figure}[htb]
\epsfysize=5cm \centerline{\epsfbox{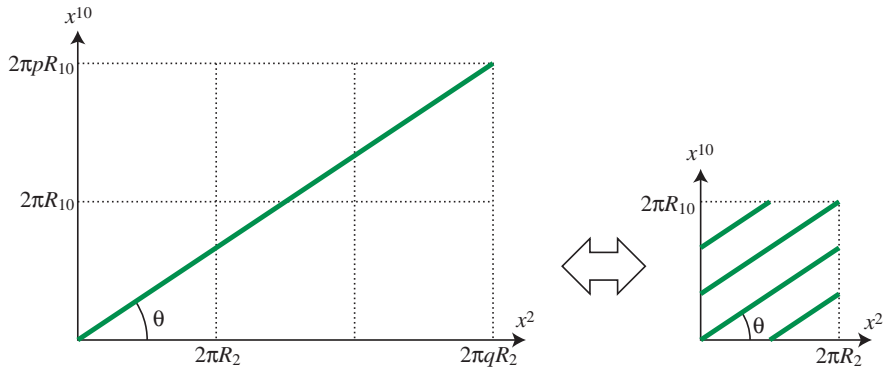}}
\caption{\small
A $(p,q)$5-brane as an M5-brane wrapping a torus $T^2$ $q$ times over $x^2$
and $p$ times over $x^{10}$.
}
\label{f1}
\end{figure}

For simplicity, we take the Type IIB string coupling equal to the
duality-invariant value $\tau_{\rm IIB}=i$. Then the relation between
the rotation angle $\theta$ and the number $(p,q)$ characterizing
the 5-brane is given by $\tan\theta=p/q$. For example, in Fig.~\ref{f1}
we depicted an M5-brane wrapping $x^2$ $q$ times and $x^{10}$ $p$
times. Obviously its rotation angle is given by $\tan\theta=p/q$. Under the
compactification and $T_2$-duality, this brane is transformed into
$(p,q)$5-brane. Indeed we find $q$ NS5-branes if we look at the brane
from $x^2$ side, and $p$ D4-branes (D5-branes) in type IIA (IIB) string
theory if we look at it from $x^{10}$.

Similarly, the M2- and M5-branes are transformed as
\bea
&& {\rm M: M2}(1|6|) \Rightarrow {\rm IIA: D2}(1|6|)
\stackrel{T_2-{\rm dual}}{\Rightarrow} {\rm IIB: D3}(12|6|), \nn
&& {\rm M: M5}(1789\natural) \Rightarrow {\rm IIA: D4}(1789)
\stackrel{T_2-{\rm dual}}{\Rightarrow} {\rm IIB: D5}(12789).
\label{adm}
\ena
We thus see that our branes without rotation precisely reproduce those
considered in ref.~\cite{HW}. Our configurations generalize the right
5-brane to a rotated $(p,q)$5-brane.

The BPS brane configurations discussed in the previous subsection thus
produce various brane configurations in string theories. In particular,
our discussions gives a systematic classification of various possible
brane configurations, including $(p,q)$5-branes hitherto not considered
except in ref.~\cite{GGPT}. The field theories realized on the common
worldvolumes of these branes are two-dimensional in type IIA theories
and three-dimensional in type IIB. We will concentrate on the latter
three-dimensional theories in this paper.

The remaining supersymmetry counted in $d=3$ dimensions are also summarized
in table~1. Our procedure also gives the exhaustive list of possible
gauge field theories on the common worldvolume. In particular, we will
see that our new configurations produce interesting gauge theories
involving CS terms for gauge fields. Our results also include $d=3$
theories with $N{=}3$ extended supersymmetry whose realization on the brane
configuration has not been disccused much. We will summarize what theories
are realized on the branes in sect.~4. Before presenting this result, let us
discuss how CS terms emerge in the brane configurations with $(p,q)$5-brane.

\section{CS terms}

The brane configurations without rotation in 11-th direction have been
discussed in ref.~\cite{HW}. In this section, we present several arguments
that the field theories realized on the common worldvolume of the branes
rotated in the 11-th direction are three-dimensional gauge theories
with CS terms.

The basic idea is very simple. Consider theories without rotation.
This is the model given in ref.~\cite{HW} and the theory is three-dimensional
gauge theory with some matters in the adjoint representation. Next,
rotate the second M5-brane or $(p,q)$5-brane to the extreme at the angle
$\theta = \frac{\pi}{2}$, resulting in D5-brane, and we find that the gauge
excitation is not allowed~\cite{HW}. This means that the gauge field
becomes extremely massive and decouples from the theory. In dimensions
other than 3, this would imply that the gauge symmetry is broken by the
explicit mass terms. However, it is possible in 3 dimensions to
give masses to the gauge fileds without breaking gauge invariance by
the CS terms~\cite{DJT}. Indeed, we will give several arguments that
this is the case in the theories on our brane configurations. Since the
matter M5-branes do not play any role in this discussion, we omit explicit
mention of these in this section. Also we discuss only the bosonic parts
since the supersymmetric generalization is straightforward.

\subsection{M5-brane solution}

We begin by considering the eleven-dimensional supergavity solutions with
an M5-brane rotated by two angles $\rot{2}{\natural}{\theta}$ and
$\rot{5}{9}{\rho}$ for simplicity:
\bea
{ds_{11}}^2 &=& H^{-1/3}\left\{
-(dx^0)^2+(dx^1)^2+(\cos\theta dx^2+\sin\theta dx^{10})^2
 + (dx^3)^2+(dx^4)^2 \right. \nn
&& \hs{15} \left. +(\cos\rho dx^5+\sin\rho dx^9)^2 \right\} \nn
&& + H^{2/3}\left\{ (dx^6)^2+(dx^7)^2+(dx^8)^2
 +(-\sin\rho dx^5+\cos\rho dx^9)^2 \right. \nn
&& \hs{15} \left. + (-\sin\theta dx^2+\cos\theta dx^{10})^2 \right\},
\label{sug}
\eea
where $H$ is a harmonic function in $(x^6,x^7,x^8,x^9,x^{10})$. This
solution does not involve the intermediate D3-branes and another NS5-brane;
these will be considered probes introduced later.

Dimensionally reducing the solution~(\ref{sug}) to Type IIA, one finds
the RR-gauge and dilaton fields are given by
\bea
A_{2} &=& \frac{\sin\theta\cos\theta(1-H)}{\sin^2\theta+H\cos^2\theta}, \nn
\phi &=& \frac{3}{4} \log H^{-1/3} (\sin^2\theta + H\cos^2\theta).
\eea
Note that those are functions of only the coordinates $x^6,x^7,x^8,x^9$.

Making $T_2$-dual from Type IIA to Type IIB~\cite{BHO}, one finds that the
metric and RR-scalar (axion) field is given by
\bea
{ds_{\rm IIB}}^2 &=& H^{1/2} (\sin^2\theta + H \cos^2 \theta)^{1/2}
 \left[ H^{-1} \left\{ -(dx^0)^2 + (dx^1)^2 + (dx^2)^2 + (dx^3)^2
 + (dx^4)^2 \right. \right. \nn
&& \hs{-10} \left.\left. + (\cos\rho dx^5+\sin\rho dx^9)^2 \right\}
 + (dx^6)^2 + (dx^7)^2 + (dx^8)^2
 + (-\sin\rho dx^5+\cos\rho dx^9)^2 \right], \nn
\chi &=& \frac{\sin\theta\cos\theta(1-H)}{\sin^2\theta+H\cos^2\theta}.
\label{axi}
\eea
We see that the Lorentz invariance in $x^2$ direction is recovered. This
is the reason why we can get three-dimensional theories even though we
started with the solution~(\ref{sug}) without Lorentz invariance.
We will argue in the next subsection that the axion fields acts as a source
to induce the CS terms in our configuration.

\begin{figure}[htb]
\epsfysize=5cm \centerline{\epsfbox{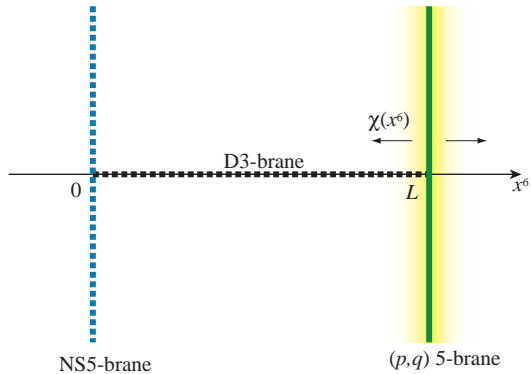}}
\caption{\small
Induced CS coupling on the D3-brane via the $(p,q)$5-brane,
which is a source of an axion field.
}
\label{f2}
\end{figure}

Here we have considered the rotated M5-brane solution and derived the axion
background. It is possible to construct the solution involving the NS5-brane
on the left, but when the $(p,q)$5-brane is far enough away from the NS5-brane,
this modification does not affect our following discussions and we can
simply disregard the complication due to its presence. To this configuration,
we also introduce D3-branes as probe as depicted in Fig.~\ref{f2}. In this
figure, the fact that the NS5- and D3-branes are introduced as probes is
indicated by their dotted lines and as such we assume that these branes do
not affect the axion background given in~(\ref{axi}). We are now going to
use this result to give a heuristic derivation of the CS terms.

\subsection{Induced CS coupling on D3-brane}

In this subsection, we discuss the derivation of CS term intuitively.
More concrete discussions will be given in sect.~3.3.

The effective action for the axion and gauge fields involves a term
\bea
\frac{1}{8\pi^2}\int_{{\cal M}^4}\chi F \wedge F,
\label{inst}
\ena
where we assume that ${\cal M}^4=\R^2\times S^1\times I$ and $I=[0,L]$.

The harmonic function is given by $H=1+\frac{l_s^2}{r^2}$,
($r \equiv \{(x^6-L)^2 +(x^7)^2 + (x^8)^2 + (x^9)^2 \}^{1/2}$),\footnote{
Here we have chosen the unit charge for the M5-brane for simplicity.}
where the position of the D3-branes is chosen at $x^7=x^8=x^9=0$. Hence the
expectation value of the axion~(\ref{axi}) has its main support only for
the range of IIB string length $l_s$. Since the length $L$ of the D3-branes
is taken large compared to the string length, the axion background has its
main value only near the $(p,q)$5-brane. The induced coupling (\ref{inst})
then implies that instantons can exist only close to the $(p,q)$5-brane
in the space of $(x^0,x^1,x^2,x^6)$.

Instanton solution represents a point in four-dimensional space. If we neglect
its size, its distribution can be approximated as
\bea
F \wedge F \sim \d^4(x) d^4 x.
\ena
Since the directions $(x^0,x^1,x^2)$ are extended infinitely, it is natural
to put this part to the three-form $AdA$, but approximate
the distribution in the $x^6$ direction by a localized delta function.
These considerations lead to the following form for the field strengths:
\bea
F \wedge F \sim \widetilde{A}\widetilde{d}\widetilde{A}
\wedge 2\pi\delta(x^6-L)dx^6,
\label{dist}
\ena
where tilde means that the quantities are independent of the coordinate
$x^6$.

Substituting eq.~(\ref{dist}) into eq.~(\ref{inst}), we get
\bea
\frac{1}{8\pi^2}\int_{{\cal M}^4}\chi(x^6) F \wedge F
&=&\frac{1}{4\pi}\int_{{\cal M}^4}\chi(x^6) \widetilde{A}\widetilde{d}
\widetilde{A} \wedge \delta(x^6-L)dx^6, \nn
&=& \frac{1}{4\pi}\chi(L)\int_{\R^2\times S^1}AdA, \nn
&=& \frac{\kappa}{4\pi}\int_{\R^2\times S^1}AdA,
\label{cs1}
\eea
where use has been made of the relation
\bea
\lim_{x^6\rightarrow L} \chi(x^6) &=& - \tan\theta \nn
&=& -p/q \equiv \kappa.
\label{kap}
\eea
In evaluating (\ref{kap}), we used eq.(\ref{axi}) and
$\lim_{x^6\rightarrow L}H=\infty$.

Thus we find that the theory contains CS terms for gauge fields.
Note that these terms are consistent with the fact that the gauge fields
are massless in the parallel configuration $(\theta=0)$ and extremely
massive in the orthogonal case $(\theta=\frac{\pi}{2})$. Since these
terms are of topological nature, we expect that the coefficients will
not have quantum corrections.

As we will discuss in sect.~5, precisely this coefficient (\ref{kap}) yields
the correct vacua for the MCS theories under study. If we look at the
brane configurations given in the right of Fig.~\ref{f1}, we notice that
the M2-branes can attach only at $p$ places, giving $p$ different vacua.
We will see that this vacuum structure is reproduced with this coefficient.
Though the above derivation is rather heuristic, we can regard this
coincidence as another evidence of our argument for the existence of
the CS terms.

\subsection{Boundary conditions and CS terms}

We now give a second argument why CS terms are induced in our theories.
For simplicity, we consider the abelian field theory in this section.

The field theories we are interested in can be regarded as realized on
the D3-branes joining the two 5-branes. First let us suppose that
we have theories on D3-branes without any restriction. The theories are
given by gauge theories in 4 dimensions with a coupling constant $g_4$. The
relevant part of the action would be
\bea
- \frac{1}{4{g_4}^2} \int d^3 x dx^6 F_{\mu\nu} F^{\mu\nu},
\label{act1}
\ena
where $d^3x \equiv dx^0 dx^1 dx^2$
and $\mu,\nu$ run over $0,1,2,6$. Actually the theories are confined to
the region $0 \leq x^6 \leq L$ by the 5-branes. What kind of boundary
conditions do these 5-branes impose on the field theories?

In the case of NS5-brane, the appropriate boundary condition is
\bea
F_{\mu 6} = \pa_\mu A_6 - \pa_6 A_\mu = 0,
\label{ns5}
\ena
where $\mu$ runs over only $0,1,2$. For D5-brane, it is
\bea
F_{\mu \nu} = \pa_\mu A_\nu - \pa_\nu A_\mu = 0.
\label{d5}
\ena
These conditions kill some of the degrees of freedom of gauge fields~\cite{HW}.
Note that the condition~(\ref{d5}) may be equivalently written as
$\e_{\mu\nu\lambda}F^{\nu \lambda} = 0$.

For the general $(p,q)$5-branes, we expect that the boundary condition
should be an $SL(2,\Z)$ combination of the conditions~(\ref{ns5}) and
(\ref{d5}). Since the NS5-brane $\left( 0 \atop{1}\right)$ is transformed
into the general $\left( p \atop{q}\right)$ 5-brane by the $SL(2,\Z)$
transformation
\bea
\left( \begin{array}{cc}
r & p \\
s & q 
\end{array}
\right),
\ena
the boundary condition is expected to be given by applying its inverse
transformation
\bea
\left( \begin{array}{cc}
q & -p \\
-s & r 
\end{array}
\right),
\ena
to the conditions~(\ref{ns5}) and (\ref{d5}). Thus we have
\bea
\pa_\mu A_6 - \pa_6 A_\mu = 0,\ \ \ {\rm at} \ \ \ x^6=0, \nn
\pa_\mu A_6 - \pa_6 A_\mu - a \frac{p}{q} \e_{\mu\nu\lambda} \pa^\nu A^\lambda
= 0,\ \ \ {\rm at} \ \ \ x^6 = L,
\label{bc}
\ena
where $a$ is a nonvanishing constant to be determined by other consideration.
Indeed, the boundary condition~(\ref{bc}) reproduces (\ref{ns5}) for
$(p,q)=(0,1)$ and (\ref{d5}) for $(p,q)=(1,0)$.

How are these conditions imposed on the theory? In field theories with
boundary, we have to be careful about the boundary terms which arise when
we make the variation of the action. When written for finite range of $x^6$,
the action~(\ref{act1}) becomes
\bea
-\frac{1}{4{g_4}^2} \int d^3 x dx^6 \left[ F_{\mu\nu} F^{\mu\nu} + 2(\pa_6 A_\mu)^2
 + 2(\pa_\mu A_6)^2 - 4 \pa_\mu A_6 \pa_6 A^\mu \right],
\label{act2}
\ena
where $\mu,\nu$ run over only $0,1,2$. We then look for the surface terms
from the boundary of $x^6$ coordinate in making the variation of the action
with respect to $A_\mu$. From the variation of the second and last terms
in eq.~(\ref{act2}), we have the surface terms
\bea
\frac{1}{{g_4}^2} \int d^3 x \d A_\mu \left[ \pa_6 A^\mu
 - \pa^\mu A_6 \right],
\label{bt}
\ena
at $x^6=0$ and $x^6=L$. These must vanish by themselves or should be
cancelled by some additional action at the boundary. In view of the boundary
conditions~(\ref{bc}), we note that the contribution from $x^6=0$ vanishes but
that from $x^6=L$ needs additional contribution from the boundary action.
It is given precisely by the variation of the CS terms
\bea
\frac{a}{{g_4}^2} \frac{p}{q} \int_{x^6=L} d^3 x \e_{\mu\nu\lambda}
 A^\mu \pa^\nu A^\lambda.
\label{cs2}
\ena
Our total action hence should be the sum of (\ref{act2}) and (\ref{cs2}).

Let us now consider the effective action for ``massless modes''.
If we were to restrict all the fields to $x^6$-independent functions,
the boundary condition~(\ref{bc}) would eliminate all degrees of
freedom~\cite{HW}. However, we would like to point out that there is
a more general possibility.

``Massless modes'' mean that if we assume factorization of the
gauge fields into $x^\mu$-dependent and $x^6$-dependent functions,
the $x^6$-dependent functions should obey Laplace equation in the $x^6$
direction, $\pa_6^2 f(x^6)=0$. A general solution for this Laplace equation
in one dimension is a linear function in $x^6$. Under this assumption,
the boundary condition~(\ref{bc}) may be satisfied by setting
$\pa_\mu A_6 = a \frac{p}{q} \frac{x^6}{L}\e_{\mu\nu\lambda}\pa^\nu
A^\lambda$ with $x^6$-independent $A_\mu$.\footnote{Apart from the
$x^6$-dependence, this is similar to the duality transformation of a vector
field into a scalar in 3 dimensions. Thus the field $A_6$ is expressed by
the ``massless modes'' in $A_\mu$.} After substituting this relation into the
action~(\ref{act2}) and integrating over the $x^6$ coordinate, we see that our
effective theory contains the usual gauge kinetic terms from (\ref{act2})
and the desired CS terms (\ref{cs2}), but the normalization of the kinetic
term is $-\frac{L}{4{g_4}^2}\left[ 1+\frac{1}{3}\left( a \frac{p}{q}\right)^2
\right]$. As we will see below, our picture is good for small $a\frac{p}{q}$
and we can drop its square in this factor. Defining the gauge coupling by
$1/g^2=L/{g_4}^2$, we find that in order to get the results consistent with
those in subsection~3.2, we must have
\bea
\frac{a}{{g_4}^2} = -\frac{1}{4\pi}.
\ena
Thus the theory we have here is indeed the gauge theory with CS terms.
The mass of the gauge field is given by
\bea
\frac{1}{2\pi} \frac{{g_4}^2}{L}\frac{p}{q}.
\ena

We may consider Kaluza-Klein massive ({\it i.e.} $x^6$-dependent) modes
for $A_\mu$ as well. In order that our above theories make sense, the
masses of these modes must be large compared to the CS mass such that
they can be considered decoupled. These modes have masses of order $1/L$
coming from the $x^6$-dependence in addition to the CS mass.
For large $\frac{p}{q} {g_4}^2$, Kaluza-Klein masses may be smaller than
the gauge-invariant CS mass and these modes may not be neglected in
the low-energy effective theories, but they can be ignored for small
$\frac{p}{q}{g_4}^2$, consistent with our above approximation. Thus our
picture is valid as long as $\frac{p}{q}{g_4}^2$ is not very large, and
of course when this is very large, all these modes are decoupled from
the theory.

A final comment on the coefficient of the CS terms is in order. For
nonabelian gauge theory, it has been argued that the CS coefficient should
be integers because of topological consideration~\cite{DJT}. We expect
that our above derivation of the CS terms is also valid for nonabelian
theory obtained for $N_c >1$ and then the question arises as to the
consistency of the theory. The topological argument is true, however,
for the spherical space-time topology, and does not apply straightforwardly
to our case of different topology $\R^2 \times S^1$. Also such restriction
does not exist for abelian theories. As discussed in refs.~\cite{EGKRS,GK},
even for nonabelian $N_c>1$ case, quantum corrections produce repulsive
force between the D3-branes, resulting in Coulomb phase in which the gauge
symmetry is broken to abelian. Thus we do not expect to encounter any quantum
inconsistency. On the contrary, we will see in sect.~5 that precisely this
fractional coefficient gives a vacuum structure consistent with brane picture.

\section{Three-dimensional field theories and their mirrors}

\subsection{Abelian field theories}

Let us now turn to the identification of the concrete field theories
realized on the D3-branes and their matter contents. We consider the case
in which the three directions of the worldvolume of $(p,q)$5-brane are
taken general, that is, the worldvolume directions $(x^3,x^4,x^5)$ are
mixed with the other directions $(x^7,x^8,x^9)$. In M-theoretic
point of view, this is the most general case of the four angles discussed
above; we can obtain the results for less angles just by setting some of the
four angles to zero. There are also the cases without the mixing of
$x^{10}$ and another direction but with rotations, for example, in $x^4$-$x^8$
or $x^5$-$x^9$. The two-angle mixings like these are also discussed in
ref.~\cite{BHO3}. We will mainly concentrate on the cases containing
the rotation of $x^{10}$ into another direction in our study of the effects
of $(p,q)$5-brane.

Let us first consider the simplest situation that one D3-brane in the
directions (126) exists between NS5 in (12345) and $(p,q)$5-brane in
$(12\rot{3}{7}{\psi} \rot{4}{8}{\varphi} \rot{5}{9}{\rho})$ as indicated in
table~\ref{t1}. By adding the $N_f$ D5-branes in (12789), we get $U(1)$ gauge
theory with $N_f$ flavors in the fundamental representation as depicted in
the left of Fig.~3.\footnote{See eqs.~(\ref{tdual}) and (\ref{adm}) for
these worldvolumes.} To clarify the main point of our study, we first
restrict ourselves to the $U(1)$ gauge theory with massless flavors and no
Fayet-Iliopoulos (FI) terms. The masses of the particles with flavors
correspond to the differences in the positions of D5-branes in (12789) from
D3-brane in the directions of $(x^3,x^4,x^5)$ and we mean by massless flavors
that these distances are zero. The FI terms, on the other hand, come from
the relative positions of the two 5-branes in the directions of $(x^7,x^8,x^9)$
and no FI terms mean that these positions of the two 5-branes are the
same.\footnote{The FI terms can be defined only when the
relative positions of the two 5-branes may be defined. As we discuss
below, when one of the two 5-branes is rotated in some directions of
$(x^7,x^8,x^9)$, the FI terms corresponding to these directions cannot
be defined.}

The matter contents can be read off from the brane configurations.
To write down the field theories, we use the three-dimensional $N=1$
superspace formalism in ref.~\cite{Iv}. We denote three-dimensional
$N{=}2$ vector multiplet superfield by $V$ and $N{=}2$ chiral superfield by
$\Phi$.\footnote{Note that the three-dimensional $N=4$ vector multiplet
consists of $V$ and $\Phi$~\cite{HW}. See ref.~\cite{ST} for the representation
theory of supersymmetry in various dimensions.}
These superfields $V$ and $\Phi$ in turn consist of three-dimensional $N=1$
superfield $U$, $X_5$ and $X_3$, $X_4$, where $U$ is an $N=1$ superfield
for the vector multiplet, and $X_3$, $X_4$ and $X_5$ are those for $N=1$ chiral
multiplets, respectively. Here our notation is that $X_3$, $X_4$
and $X_5$ come from the fluctuations of the D3-brane in the directions of
$x^3$, $x^4$ and $x^5$. We also denote $N_f$ (anti-)fundamental matters
as $Q$ ($\tilde Q$) which are the $N{=}2$ chiral superfields.  They
originate from the strings stretching between $N_f$ D5-branes in (12789)
and the D3-brane.

Using these notations, we can write down the actions for the brane
configurations. In terms of $V$ (also its super field strength $W$),
$\Phi$ and $Q$($\tilde Q$) in $N=2$ superspace, we have the well-known
three-dimensional $N=4$ actions for supersymmetric $U(1)$ gauge theory
with $N_f$ flavors~\cite{HW}:
\bea
S_{N{=}4}&=& \frac{1}{g^2} 
\left[ \int d^3 x d^4 \theta \Phi^{\dagger}\Phi + \frac{1}{2}
   \left( \int d^3 x d^2 \theta W^2 + {\rm h.c.}\right) 
   \right] \nonumber\\ 
&&+ \int d^3 x d^4 \theta \left(Q^{\dagger}e^{2V}Q
 + \tilde Q e^{-2V} \tilde Q^{\dagger} \right)
+ \frac{1}{\sqrt{2}}\left( \int d^3 x d^2 \theta \tilde Q \Phi Q
 + {\rm h.c.}\right).
\eea
In addition, we need the actions which generally reduce the
$N=4$ $U(1)$ field theory to that with $N=1$:
\be
S=\int d^3x d^2 \theta \frac{-i}{4} \Big[ \kappa_0
(D^{\gamma} U^{\beta} D_{\beta} U_{\gamma} ) - \sum_{l=3,4,5}
\kappa_l X_l^2 \Big],
\label{ele}
\ee
where $D^\c$ and $\theta$ are the covariant derivative and fermionic
super-coordinates in three-dimensional $N{=}1$ superspace, respectively,
and $\b$ and $\c$ are the indices of three-dimensional spinor~\cite{Iv}. 
In eq.~(\ref{ele}), $\kappa_l$ ($l$=0,3,4,5) are coupling constants, which
represent CS terms for gauge fields and masses for the scalars $X_l$
($l=$3,4,5). These actions show the effects of the $(p,q)$5-brane, rotated
in the directions of $x^5$-$x^9$, $x^4$-$x^8$ and $x^3$-$x^7$ relative to the
NS5-brane in (12345), are the production of mass terms for the theories
on the branes. Since there are certain relations among these rotation
angles to ensure partially
unbroken supersymmetry, we expect that there are also some relations among
these masses $\kappa_l$ ($l$=0,3,4,5). We will come back to these conditions
after discussing the magnetic theory.
\begin{figure}[htb]
\epsfysize=5cm \centerline{\epsfbox{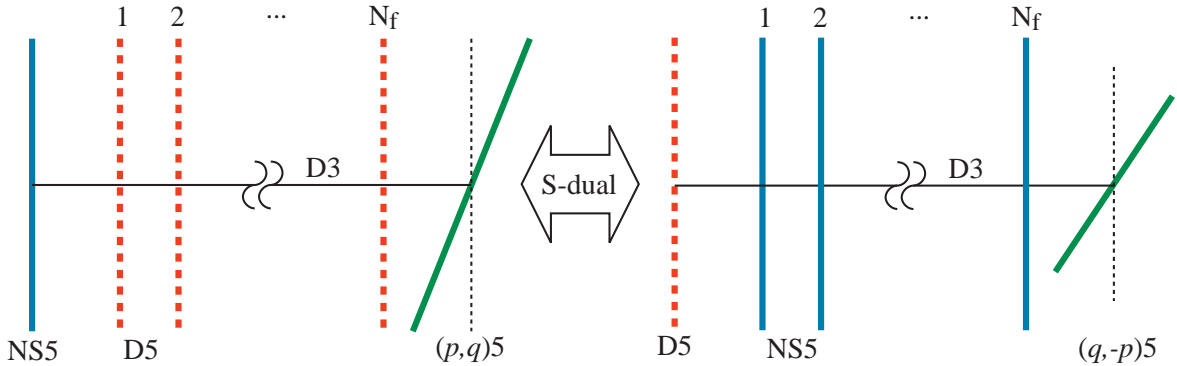}}
\caption{\small
CS theory with $N_f$ matters and its dual theory.}
\label{dual0}
\end{figure}

\subsection{Mirrors of abelian field theories}

The mirror theories of the above original theories can be obtained by the
S-transformation, which is contained in the known $SL(2,\Z)$ symmetry of
type IIB string theory (see Fig.~3). We call these mirror theories the
magnetic ones as opposed to the original theories which are called the electric
ones. By this transformation, the above brane configurations change into
those in which one D3-brane in (126) is suspended between D5 in (12345) and
$(q,-p)$5-brane rotated as $(12\rot{3}{7}{\psi} \rot{4}{8}{\varphi}
\rot{5}{9}{\rho})$. The $N_f$ D5-branes also change into $N_f$ NS5-branes.
As a result, the D3-brane between D5-brane and $(q,-p)$5-brane is separated
into the $N_f$ +1 segments by $N_f$ NS5-branes, so that we have
three-dimensional pure $U(1)^{N_f -1} \times U(1)$ gauge theory with CS
coupling only in the last $U(1)$ theory.\footnote{We neglect the field theory
on the part of D3-brane between the left D5-brane and the NS5-brane next to
the D5-brane (in other words, the 0-th segment), because the matters on
this segment are considered very massive.}  The last $U(1)$ factor with CS
coupling comes from the NS5-brane nearest to the $(q,-p)$5-brane. This part
plays an important role in reducing $N=4$ supersymmetry to the lower
supersymmetry as we discuss below.

We denote three-dimensional $N{=}2$ vector and chiral multiplets for the
$i$-th $U(1)$ factor, ($i$=0,1...$N_f$) by $V^{(i)}$ and $\tilde \Phi_{(i)}$,
respectively. The superfields $V^{(i)}$ consist of three-dimensional
$N=1$ vector multiplet $U^{(i)}$ and $N=1$ chiral multiplet $X_9^{(i)}$.
The fields $\tilde \Phi_i$,
on the other hand, consist of three-dimensional $N=1$ chiral multiplets
$X_8^{(i)}$ and $X_7^{(i)}$. As in the case of the electric theory,
$X_7^{(i)}$, $X_8^{(i)}$ and $X_9^{(i)}$ come from the fluctuations of
the D3-branes in the directions of $x^7$, $x^8$ and $x^9$. In addition to
them, there are also bi-fundamental matters
$b_{i,i+1}$($\tilde b_{i,i+1}$) which transform in the (anti-)fundamental
representation for each $U(1) \times U(1)$ on the neighboring segments of
the D3-branes. These matters come from the strings stretching between the
neighboring segments.

The actions for the magnetic theories also consist of the $N=4$ part and the
non-$N=4$ part. The $N=4$ part is $U(1)^{N_f}$ gauge
theory which is the dual of the $N=4$ action of the above
electric theory~\cite{HW}:
\bea
S_{N=4}&=& \sum_{i=1}^{N_f} \Big[
     \frac{1}{g_{(i)}^2}  \Big(
    \int d^3 x d^4 \theta \Phi_{(i)}^{\dagger}\Phi_{(i)} + 
    \frac{1}{2} ( \int d^3 x d^2 \theta W_{(i)}^2 + {\rm h.c.}) 
                 \Big)   \nn
&+& \int d^3 x d^4 \theta \Big(
       b_{i-1,i}^{\dagger}(e^{2V_{(i)}}+ e^{-2V_{(i-1)}}) b_{i-1,i}
       +  \tilde b_{i-1,i}^{\dagger}(e^{2V_{(i-1)}}+ e^{-2V_{(i)}}) 
     \tilde b_{i-1,i}    \Big)  \nn
&+&  \frac{1}{\sqrt{2}} \Big( 
 \int d^3 x d^2 \theta  b_{i-1,i}(\Phi_{(i)} - \Phi_{(i-1)}) \tilde b_{i-1,i}
    + {\rm h.c.} \Big)  \Big],
\eea
where $g_{(i)}$ is the gauge coupling of the $i$-th $U(1)$ factor and we have
defined that $V_{(0)}$ and $\Phi_{(0)}$ are zero. On the other hand, the
non-$N=4$ part is
\be
S=\int d^3x d^2 \theta
\frac{-i}{4} \Big[ \tilde \kappa_0 (D^{\gamma} U_{(N_f)}^{\beta}
D_{\beta} U^{(N_f)}_{\gamma} ) - \sum_{l=7,8,9} \tilde \kappa_l
X^{l2}_{(N_f)} \Big],
\label{mag} 
\ee
where we have defined the $N_f$-th $U(1)$ gauge theory by that on the segment
between $(q,-p)$5-brane and the NS5-brane nearest to the $(q,-p)$5-brane.
As in the case of electric theories, $\tilde \kappa_l$ ($l$=0,7,8,9) represent
the masses for the gauge field and the scalars of the $N_f$-th $U(1)$ gauge
theory. The two brane-configurations related by S-duality show that the
vacua or moduli of one of the two different field theories correspond to those
of the other. From the brane configurations, we can see that the Higgs branch
of the electric theories corresponds to the Coulomb branch of the other, and
vice versa. This is reminiscent of $N=4$ electric-magnetic duality~\cite{HW}.

\subsection{Conditions for couplings and unbroken supersymmetry}

Up to this point, we only give the general form of the actions for abelian
theories on the branes. The actual theories, of course, have particular
masses and couplings, depending on the remaining supersymmetry.
Let us now present the conditions imposed by the unbroken supersymmetry
for the above brane configuration.

Using the same notation as in table~\ref{t1}, we sum up the
results in table~\ref{t2} below. We only use the fact that the relative angles
for the two 5-branes lead to the mass terms in the field theories on the
D3-branes so as to be consistent with the remaining supersymmetry.
We have excluded the freedom of the trace part of the mass matrix of flavors
from the mass parameters because this freedom is absorbed into the
vacuum expectation values of the adjoint scalars. For completeness,
in addition to the results for configurations involving
the $(p,q)$5-brane, we also add the cases of two NS5-branes rotated in
various directions. In the results summarized in this table, we also
assume that all the angles are not $\frac{\pi}{2}$.

\begin{table}[htb]
\begin{tabular}{|c|c|c|c|c|}
\hline
 & coupling & massless adj. & mass param. & FI param.\\
\hline
\hline
1      & $\kappa_l=0$ ($l$=0,3,4,5) & 3 & $3(N_f-1)$ & 3 \\
\hline
2-(i)  & $\kappa_4= \kappa_5$, $\kappa_0= \kappa_3=0$ &1 & $3(N_f-1)$ & 1 \\
\hline
2-(ii) & $\kappa_0= \kappa_5$, $\kappa_4= \kappa_3=0$ &2 & $3(N_f-1)$ & 2 \\
\hline
3-(i)  & $\kappa_0=0$,  $f^{(1)}(\kappa_3,\kappa_4, \kappa_5)=0$ & 0 & 
$3(N_f-1)$ & 0 \\
\hline
3-(ii) & $\kappa_3=0$,  $f^{(2)}(\kappa_0,\kappa_4, \kappa_5)=0$ &1 & 
$3(N_f-1)$ & 1 \\
\hline
4-(i)  & $f^{(3)}(\kappa_0,\kappa_3,\kappa_4, \kappa_5)=0$ & 0 &
$3(N_f-1)$ & 0\\
\hline
4-(ii) & $\kappa_3= \kappa_4$, $\kappa_0= \kappa_5$  & 0 & $3(N_f-1)$ & 0\\
\hline
4-(iii)& $\kappa_0= \kappa_5 =\kappa_4= \kappa_3 $ &0 & $3(N_f-1)$ & 0\\
\hline
\end{tabular}
\caption{\small
Relations among angles and couplings, and other parameters for various
supersymmetry in the electric theories.
}
\label{t2}
\end{table}

The functions $f^{(i)}$, ($i=1,2,3$) in this table mean that there are certain
constraints on the masses $\kappa_l$ coming from the conditions on
the angles for unbroken supersymmetry. Only with the present information at
hand, it is difficult to determine the exact forms of these functions, but we
know for sure that there must exist such functions in order to match the
numbers of free parameters of our choice.

Similarly there are some relations for the masses in the magnetic theories.
The results are obtained in the similar way to the electric theories and
are given in table~\ref{t3}. The unknown functions $\tilde f^{(i)}$ again
originate from the constraints on the angles for unbroken supersymmetry
as in the case of the electric theories. The infinite masses in this table
imply that these degrees of freedom are very heavy and decouple from
the theory. The essential part in the magnetic brane configurations in
reducing $N=4$ SUSY to the lower supersymmetry is the segment of D3-branes
between $N_f$-th NS5-brane in (12789) and the tilted $(q,-p)$5-brane.
The above conditions for the couplings coming from this part
of the brane configurations and the other parts are the same as in the case
of $N=4$ $U(1)^{N_f -1}$ gauge theories~\cite{HW}.

\begin{table}[htb]
\begin{tabular}{|c|c|c|c|c|}
\hline
 & coupling & massless adj. & mass param. & FI param.\\
\hline
\hline
1      & $\tilde \kappa_l= \infty$ ($l$=0,7,8,9) & $3(N_f-1)$ & $3$ 
& $3(N_f-1)$ \\
\hline
2-(i)  & $\tilde \kappa_8= \tilde \kappa_9$, $\tilde \kappa_0= \tilde
\kappa_7=\infty$& $3(N_f-1)$ & $1$ & $3(N_f-1)$ \\
\hline
2-(ii) & $\tilde \kappa_0= \tilde \kappa_9$, $\tilde \kappa_8= \tilde
\kappa_7=0$ & $3(N_f-1)$ & $2$ & $3(N_f-1)$ \\
\hline
3-(i)  & $\tilde \kappa_0= \infty$, $\tilde f^{(1)}
(\tilde \kappa_7,\tilde \kappa_8, \tilde \kappa_9)=0$& $3(N_f-1)$ & 
$0$ & $3(N_f-1)$ \\
\hline
3-(ii) & $\tilde \kappa_7=\infty$,  $\tilde f^{(2)}
(\tilde \kappa_0,\tilde \kappa_8, \tilde \kappa_9)=0$ & $3(N_f-1)$ & 
$1$ & $3(N_f-1)$ \\
\hline
4-(i)  & $\tilde f^{(3)}(\tilde \kappa_0,\tilde \kappa_7,\tilde \kappa_8,
\tilde \kappa_9)=0$ & $3(N_f-1)$ & $0$ & $3(N_f-1)$ \\
\hline
4-(ii) & $\tilde \kappa_7= \tilde \kappa_8$,
 $\tilde \kappa_0= \tilde \kappa_9$ & $3(N_f-1)$ & $0$ & $3(N_f-1)$\\
\hline
4-(iii)& $\tilde \kappa_0= \tilde \kappa_9 =\tilde \kappa_8
 = \tilde \kappa_7$ &$3(N_f-1)$ & $0$ & $3(N_f-1)$\\
\hline
\end{tabular}
\caption{\small
Relations among angles and couplings, and other parameters for various
supersymmetry in the magnetic theories.
}
\label{t3}
\end{table}

\subsection{Generalization}

We briefly discuss the more general cases than the above $U(1)$ gauge theories.

First of all, let us consider the abelian gauge theories with non-zero masses
for flavors and FI terms added. Again the effect of these modifications is
the same as in the $N{=}4$ field theories discussed in ref.~\cite{HW}. We can
just add the $N{=}4$ invariant mass terms and FI terms (corresponding to the
directions in which we can define relative positions of the two 5-branes) to
the above $N{=}4$ invariant actions~\cite{HW}. These additional terms also
produce the corresponding new terms in the $N{=}4$ invariant actions of the
magnetic theories. On the magnetic side, the FI parameters in the electric
theories correspond to the masses of the bi-fundamental matter coming from
the strings stretching between the two segments of D3-branes separated by
the $N_f$-th NS5-brane. The relative differences in the directions of
$x^3,x^4,x^5$ between the neighboring D5-branes in the electric theories,
that is, the differences of the flavor masses, correspond to the FI terms
of each $U(1)$ factors in the magnetic theories.

Note that the brane configuration to the left of the $(N_f-1)$-th D5-brane
is basically the same as the $N{=}4$ Hanany-Witten configuration. Hence
the correspondence of the parameters between the electric and magnetic
theories may be understood as a reflection of the original $N{=}4$
supersymmetry and it may be no wonder that there is such correspondence in
the theory with lower supersymmetry because of the remaining `$N{=}4$ part'
in our brane configurations. We remark, however, that our results for the
conditions of couplings and supersymmetry are not changed by these additional
terms.

Next, we discuss the nonabelian case in which $N_c$ D3-branes are stretched
between the two 5-branes. In this case, the action~(\ref{ele}) must
be generalized as
\bea
S&=&\int d^3x d^2 \theta \frac{-i}{4} {\rm Tr}\Big[ 
  \kappa_0 \Big(
   D^{\gamma} U^{\beta} D_{\beta} U_{\gamma}
 + \frac{2}{3} g D_{\beta} U_{\gamma} \{ U^{\beta}, U^{\gamma} \} \nn
&& + \frac{1}{6}  g^2 \{ U_{\beta}, U_{\gamma} \} \{ U^{\beta}, U^{\gamma} \}
\Big)  - \sum_{l=3,4,5} \kappa_l X_l^2 \Big],
\eea
where $\{$ $\}$ means anti-commutator, and $g$ is the gauge coupling
constant~\cite{Iv}.  Of course, $N{=}4$ actions must also be generalized to
the nonabelian theories. Our results for the conditions of couplings and
supersymmetry are the same as those of the abelian theories.

However, there is a very special situation that may be worth mentioning.
This is the case of 4-angle with $N{=}3$ supersymmetry. Here we find that
the number of D3-branes is constrained by the s-rule~\cite{HW}.
This rule states that more than one D3-brane cannot exist at the same positions
of the 5-branes, and the $(p,q)$5-brane in Fig.~\ref{f1} suggests that
the maximal number of D3-branes which can exist between the NS5- and
$(p,q)$5-branes is $p$ because the left NS5-brane exists only in the space
$x^{10}=0$.\footnote{A similar discussion of positions to which the D3-branes
can attach will be given in sect.~5.2.} Therefore the gauge symmetry is the
direct product of at most $p$ $U(1)$'s in our configuration. Beyond $p$,
the brane picture suggests that there is no supersymmetric configuration.
It would be interesting to understand how this occurs in the effective field
theory. The actions of the magnetic theories are complicated because there
are many segments of D3-branes and various matters, but they are
essentially the same as the magnetic theories of the abelian theory.

All these considerations are at the classical level. As mentioned at the end
of sect.~3, repulsive forces between D3-branes are produced by quantum
corrections and nonabelian gauge symmetries are expected to be broken
to abelian~\cite{EGKRS,GK}. These can be, of course, described by the actions
we have given here.

\section{Mirror symmetry}

In this section we investigate the relation between dualities of
CS theories and $SL(2,\Z)$ duality of Type IIB superstring
theory in detail. We will give a simplest example of ``mirror''
symmetry of the CS theory and its explanation via the brane
configuration with $(p,q)$5-brane. We will also discuss another
construction of ``mirror'' pairs of the CS theory, which
includes operations of exchanging the 5-branes. We will encounter some
new results, which was not exposed in the Hanany-Witten configurations.

\subsection{Duality between MCS theory and self-dual model}

We first consider the duality between MCS theory and
self-dual model (SDM) and the realization of this duality in Type IIB
superstring theory. The MCS theory is a three-dimensional abelian
gauge theory which has the Maxwell gauge field action and gauge-invariant
CS terms. The SDM is proposed in \cite{TPN}. This SDM also has
CS terms, but is not gauge invariant due to a mass term for
the vector field. These theories apparently look different.
However, both theories describe a single massive excitation in 3
dimensions and violate reflection symmetry. In fact, it is shown
in~\cite{DJ} that by a Legendre transformation the SDM is equivalent
to the MCS theory. Subsequently this equivalence has been investigated
by many authors using various methods~\cite{K,BRR,AR,PS}.

\begin{figure}[htb]
\epsfysize=5cm \centerline{\epsfbox{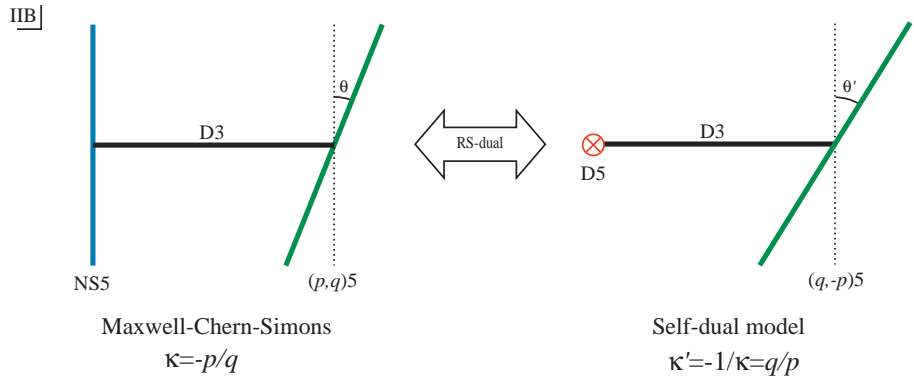}}
\caption{\small
Duality between MCS theory and SDM and their brane configurations.}
\label{dual1}
\end{figure}

We would like to realize this duality on the branes in Type IIB
theory. As we have mentioned in sect.~3, the MCS can be described on the
D3-brane between the NS5-brane and the $(p,q)$5-brane. Since the MCS
theory is an abelian gauge theory, only one D3-brane is stretched
between the NS5-brane and $(p,q)$5-brane (see the left side of
Fig.\ref{dual1}). The left NS5-brane does not
kill the degrees of freedom of the vector field and the Maxwell action
remains on the D3-brane. The right $(p,q)$5-brane will provide CS terms
with a rational coefficient $\kappa=-p/q$. So, we obtain the MCS theory
on the D3-brane in the above configuration and the action is as follows:
\bea
S_{\rm MCS}=\int d^3x \left(
\frac{\kappa}{4\pi}\epsilon^{\mu\nu\lambda}A_\mu\del_\nu A_\lambda
-\frac{1}{4g^2}F_{\mu\nu}F^{\mu\nu} \right).
\label{mcs}
\ena

Now we take an S-transformation of $SL(2,\Z)$ symmetry in Type IIB theory
and flip the coordinates $(x^3,x^4,x^5)$ and $(x^7,x^8,x^9)$
simultaneously. This operation is called the RS duality
transformation in \cite{HW} and maps a brane configuration to one of
``mirror'' or ``magnetic'' theories. This RS duality transformation is
simple from the M-theory point of view; it is only a flip
of the coordinates $(x^2,x^3,x^4,x^5)$ and $(x^{10},x^7,x^8,x^9)$.
After this RS duality transformation the configuration of the MCS
theory is as follows. The left NS5-brane is turned into the
D5-brane, whose worldvolume is $(12789)$. The right
$(p,q)$5-brane is mapped to $(q,-p)$5-brane by the S-duality. This means
that the angle $\theta$ in the $(x^2,x^{10})$-plane is changed to
$\theta'=\pi/2-\theta$. The angle in the $(x^5,x^9)$-plane must also be
$\theta'$ in order to preserve the supersymmetry. This is done by the
R-flip of the coordinates. The D3-brane is not changed and stretched between
the D5-brane and $(q,-p)$5-brane, since D3-brane is self-dual
by the RS transformation. This configuration is depicted in the right
side of Fig.~\ref{dual1}.

What is the theory on the D3-brane in the RS dual configuration? Due to
the boundary condition of the left D5-brane, zero modes of the
three-dimensional vector field will vanish. Only a mass term of the
vector field, which is due to the Higgs mechanism by the left matter
D5-brane,  remains. The right $(q,-p)$5-brane provides CS terms with
a rational coefficient $\kappa'=-1/\kappa=q/p$.\footnote{The physical
meaning of this sign change is discussed in ref.~\cite{KL2}.}
So we obtain the effective action on the D3-brane:
\be
S_{\rm SDM}=\int d^3x \left(
-\frac{4\pi}{\kappa}\epsilon^{\mu\nu\lambda}B_\mu\del_\nu B_\lambda
+2 g^2 B_{\mu}B^{\mu} \right),
\label{SDM action}
\ee
where $B_\mu$ is a massive vector field. This is nothing but the SDM
action. Thus the MCS theory and the SDM are related with
each other by the RS-duality in string theory.

In the low-energy limit of the MCS theory, the Maxwell kinetic term is
decoupled from the MCS action. So, the pure abelian CS theory, which
has only the CS terms, appears as the low-energy theory of
the MCS theory and the vacuum structure of the MCS is described by the
full Hilbert space of the pure CS theory.
On the other hand, the SDM action (\ref{SDM action}) implies
that there exists no nontrivial vacuum in the SDM and the vacuum structures
of the SDM and the MCS seem to be different.

This fact is not a contradiction for the equivalence of the MCS theory
and SDM. According to careful field-theoretical analyses, the partition
functions of the MCS theory $Z_{\rm MCS}$ and the SDM are equivalent modulo
the partition function of the pure CS theory $Z_{\rm CS}$ as~\cite{K,CKS,AR,PS}
\be
Z_{\rm MCS}(\kappa)=Z^{dual}_{\rm SDM}(-1/\kappa),
\label{partition function}
\ee
where $Z^{dual}_{\rm SDM}(-1/\kappa)=Z_{\rm CS}(\kappa)Z_{\rm
SDM}(-1/\kappa)$ and $Z_{\rm SDM}$ is a trivial vacuum sector of the SDM.
Notice that the coefficient of the CS terms in the SDM is an inverse of
the coefficient in MCS theory. The SDM dual to the MCS theory
contains non-local interactions representing the above pure CS theory, but
we simply refer to it as SDM. It follows that the exact dual of the
MCS theory has the same vacua as the pure CS theory with a CS coefficient
$\kappa$. In order to investigate the vacuum structure of both the MCS and
SDM theories, it is sufficient to consider the vacua of the pure CS theory.

{}From a field theoretical point of view, this equivalence of vacuum
structure is highly nontrivial, but we can easily understand vacuum
structures via Type IIB or M-theory branes. We will see this in the next
subsection.

\subsection{Vacuum structure of pure CS theory}

The MCS theory and its dual SDM have the same vacuum structure as the
pure CS theory. In this subsection we compare the vacuum structure of
the pure CS theory with the brane configurations in Type IIB theory or
M-theory. This can be done by adopting the field theoretical analysis of
the CS system given in ref.~\cite{IL}. The following discussions are
basically those given there suitably modified to fit our situation.

In this paper, our Type IIB brane configuration is constructed from
that of M5-branes and M2-branes in M-theory. In that process
we take T-duality once with respect to the $x^2$-coordinate, which
is compactified on $S^1$. So, we must consider field
theories in $(x^0,x^1,x^2)$-space of the D3-brane worldvolume which
is topologically $\R^2\times S^1$. This topology on which we are
considering the CS theory is crucial to the vacuum structure
discussed later.

The pure CS theory is given by
\be
S_{\rm CS}=\frac{\kappa}{4\pi}\int d^3x\epsilon^{\mu\nu\lambda}A_\mu
\del_\nu A_\lambda,
\label{CS action}
\ee
where $\kappa=-p/q$ as explained in sect. 3.
On $\R^2\times S^1$ arbitrary gauge transformations are not admitted for
an invariance of the CS action. In fact, under a gauge transformation
\be
A_\mu \rightarrow A_\mu + \del_\mu\Lambda,
\label{gt}
\ee
the pure CS action in (\ref{CS action}) changes by a surface term
\be
\delta S_{\rm CS}=\frac{\kappa}{2\pi}\int dx^0dx^2 \left.\Lambda
F_{02}\right|_{x^1=\pm\infty}.
\ee
This implies that the gauge transformation function $\Lambda$ must vanish
at infinity in $x^1$-direction for nonvanishing electric field $F_{02}$.
A periodicity in the coordinate $x^2$
(compactified on $S^1$) also requires that the function $\Lambda$ be periodic
in the $x^2$-coordinate. Henceforth we only consider this restricted class
of gauge transformations.

We rewrite the action~(\ref{CS action}) as
\bea
S_{\rm CS}=\frac{\kappa}{4\pi}\int d^3x
\left(
A_2\dot{A}_1-A_1\dot{A}_2+2A_0F_{12}
\right).
\ena
Now $A_0$ is seen to be an auxiliary field and so the variation with
respect to $A_0$ gives the constraint
\bea
F_{12}=\del_1A_2-\del_2A_1=0.
\label{F12=0}
\ena
We can write the solutions of the constraint (\ref{F12=0}) as
\bea
\mb{A}(t,\mb{x})=\mb{\nabla}U+\mb{a}(t),
\label{solution}
\ena
where $U$ is a function on the space $(x^1,x^2)$ of the topology
$\R\times S^1$, and $a_{1,2}$ are $\mb{x}$-independent variables,
which coincide with the unexponentiated Wilson line operators along
the $x^1$ and $x^2$ coordinates:
\be
a_1(t)=\int_{\R} dx^1 A_1, \qquad a_2(t)=\int_{S^1} dx^2 A_2.
\ee

Substituting (\ref{solution}) into the action (\ref{CS action}), we
obtain the effective action for the topological sector
\be
S_{top}=\frac{\kappa}{2\pi}\int dt a_2\dot{a}_1.
\ee
This action tells us that $a_1$ and $a_2$ are canonical conjugates and
satisfy the commutation relation
\be
[a_1,a_2]=2\pi i/\kappa.
\label{comm rel}
\ee

We now define the Wilson line operators
\be
W_1=\exp(ia_1), \qquad W_2=\exp(-ia_2).
\ee
These operators are gauge invariant under the above restricted
transformations~(\ref{gt}). These represent the only physical degrees of
freedom we are interested in. Using the commutation relations~(\ref{comm rel}),
we can derive those for the Wilson line operators
\be
W_1W_2=e^{2\pi i/\kappa}W_2W_1.
\label{algebra}
\ee

Recall that we are now considering the case of a rational coupling constant
$\kappa=-p/q$. The structure of Hilbert space of the Wilson line operators
is a little bit complicated in this situation. We would like to construct
the irreducible representations of the algebra (\ref{algebra}).

Let us look for a suitable Casimir operator. A candidate is $(W_2)^p$,
which commutes with $W_1$. Namely $(W_2)^p$ belongs to a center and can be
represented by a $c$-number. Since the Wilson line operator is
$U(1)$-valued, this $c$-number is also $U(1)$-valued, that is, it can be
written as an arbitrary constant phase
\be
(W_2)^p=e^{i\eta}.
\label{c-number}
\ee
This phase plays the same role as the $\theta$-vacuum angle in QCD.

To find the structure of the Hilbert space, it is convenient to
diagonalize the operator $W_2$. From the observation given in
(\ref{c-number}), we find that the possible vevs of $W_2$ are
\be
\left\langle W_2 \right\rangle = e^{i\eta/p}\omega^n,
\ee
where
\bea
&&0 \leq \eta/p < 2\pi/p,\nn
&&\omega^p=1, \quad 1\leq n \leq p,\quad n\in \Z.
\eea
This means that the pure CS theory has $p$ discrete vacua.

{}From a string theoretical point of view, the vevs of Wilson line operators
of the effective theory on the D-brane represent the positions of the D-branes
after a $T$-duality transformation~\cite{POL}. In our case, the vevs of
the Wilson line operators on the D3-brane in Type IIB theory correspond
to the positions of the D2-brane in Type IIA theory via $T_2$-dual, which
are equivalent to the positions of the M2-brane in M-theory. In this way,
in M-theory, the positions of the M2-brane can be
read from the vevs of the Wilson line operators on the D3-brane as
\be
x^2=R_2(\eta/p + 2\pi n/p),
\label{M2 position}
\ee
where $R_2$ is a radius of a compactified circle in the $x^2$-direction.
This means that the M2-brane cannot sit nowhere but $p$ discrete positions.

\begin{figure}[htb]
\epsfysize=4cm \centerline{\epsfbox{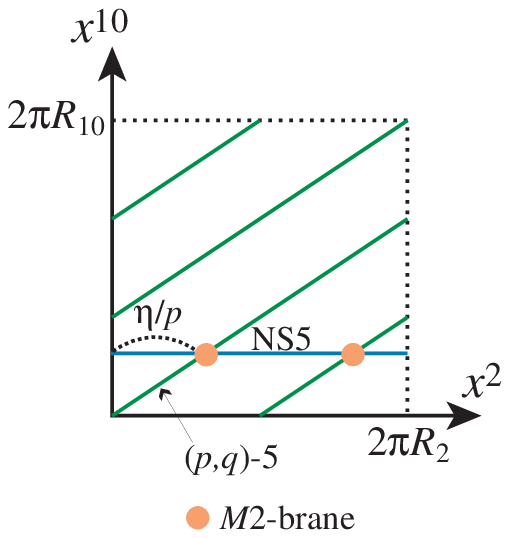}}
\epsfysize=5cm \centerline{\epsfbox{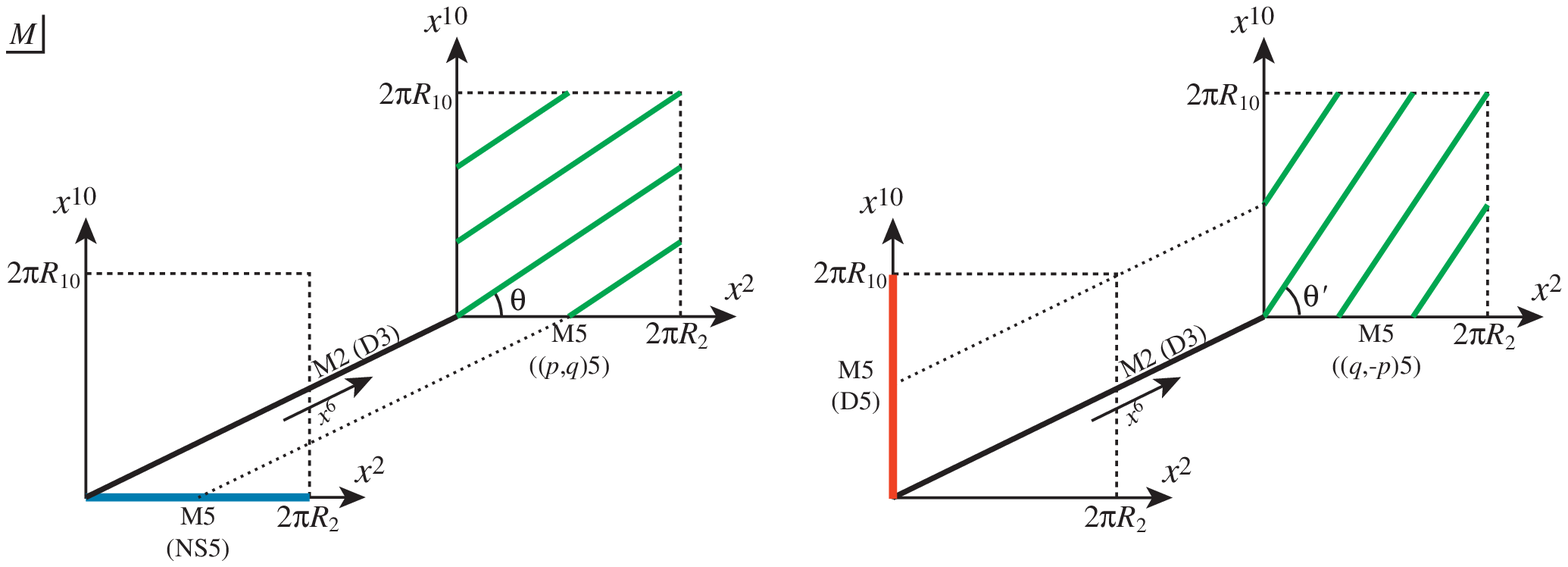}}
\caption{\small
Relation between discrete vacua of CS theory and positions
of $M$2-brane in the configuration.
}
\label{M2pos}
\end{figure}

Can we explain this phenomenon in our brane configuration? As we have
mentioned in sect.~2, the right $(p,q)$5-brane is represented as an M5-brane
wrapping on a torus. On the other hand, the left NS5-brane is an M5-brane
wrapping along the $x^2$-direction in M-theory. The M2-brane connects between
these two types of M5-branes. Since on that torus the $(p,q)$5-brane
intersects $p$ times the $x^2$-direction, along which the left NS5-brane
exists, the M2-brane can only attach to $p$ discrete positions in the
$x^2$-direction (see Fig.~\ref{M2pos}).

Moreover, it is not necessary that the NS5-brane lies on the origin of
$x^{10}$-axis. The NS5-brane can move freely in the $x^{10}$-direction. If the
$x^{10}$ position of the NS5-brane changes from $0$ to $2\pi R_{10}$, the
positions at which the M2-brane can be attached shift from $0$ to $2\pi
R_{2}/p$ simultaneously. This shift corresponds precisely to the value of the
constant term $R_2\eta/p$ in (\ref{M2 position}).
Therefore, the possible positions of the M2-brane in our configuration coincide
exactly with the values in (\ref{M2 position}), which has been derived from
the field theoretical consideration. Thus the MCS theory~(\ref{mcs}) we have
identified has the vacuum structure consisting of $p$ different vacua,
which can be easily understood in the brane picture.

On the other hand, the RS-dual configuration, which represents the
SDM, is obtained just by a coordinate flip in the M-theory configuration
space. The number of the admitted positions of M2-brane is exactly the same
as that in the electric MCS theory. Namely there are also $p$
discrete vacua in the SDM. This agrees with the
relation in (\ref{partition function}).

\subsection{More on mirrors and brane creation}

So far, we have treated a class of the ``mirror'' theories which are
constructed by the
RS-dual only. Now we will consider another example of the ``mirror'' theories
which is obtained by the operations including exchanges of the branes in
addition to the RS-dual. These exchanges of branes provide us with a more
complicated example of ``mirror'' theories.

When one exchanges the positions of two particular types of branes, another
new brane is created between the branes, that is known as the Hanany-Witten
effect~\cite{HW}. Hanany and Witten have discussed that if a NS5-brane with a
worldvolume $(12345)$ and a D5-brane with a worldvolume $(12789)$ intersect
at an $x^6$-position, then a new D3-brane with a worldvolume $(12|6|)$ is
created. In M-theory, this phenomenon can be understood as an M2-brane creation
with a worldvolume $(1|6|)$ by the crossing of an M5-brane with a worldvolume
$(12345)$ and an M5-brane with a worldvolume $(1789\natural)$.

In our brane configuration, $(p,q)$5-brane is included in contrast to the
Hanany-Witten configuration. We must consider the brane creation in the crossing
process of the NS5- or D5-branes and the $(p,q)$5-brane. The NS5-, D5- and
$(p,q)$5-branes in Type IIB theory are all represented by an M5-brane
in the M-theory. Thus in the M-theory all types of the brane creations in
the 5-brane crossing can be understood as a same single phenomenon as
the Hanany-Witten case, that is, the M2-brane creation by the M5-branes.

\begin{figure}[htb]
\epsfysize=4cm \centerline{\epsfbox{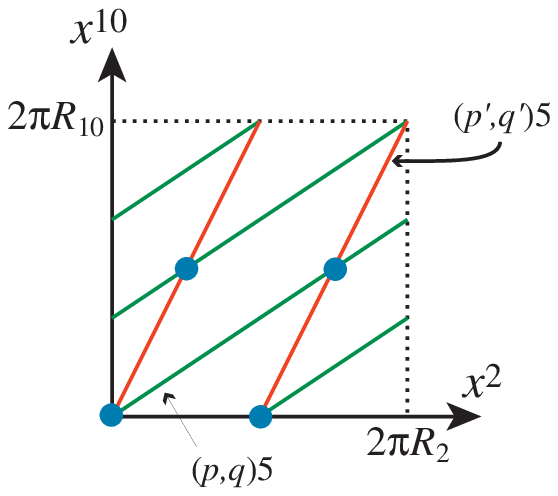}}
\epsfysize=4cm \centerline{\epsfbox{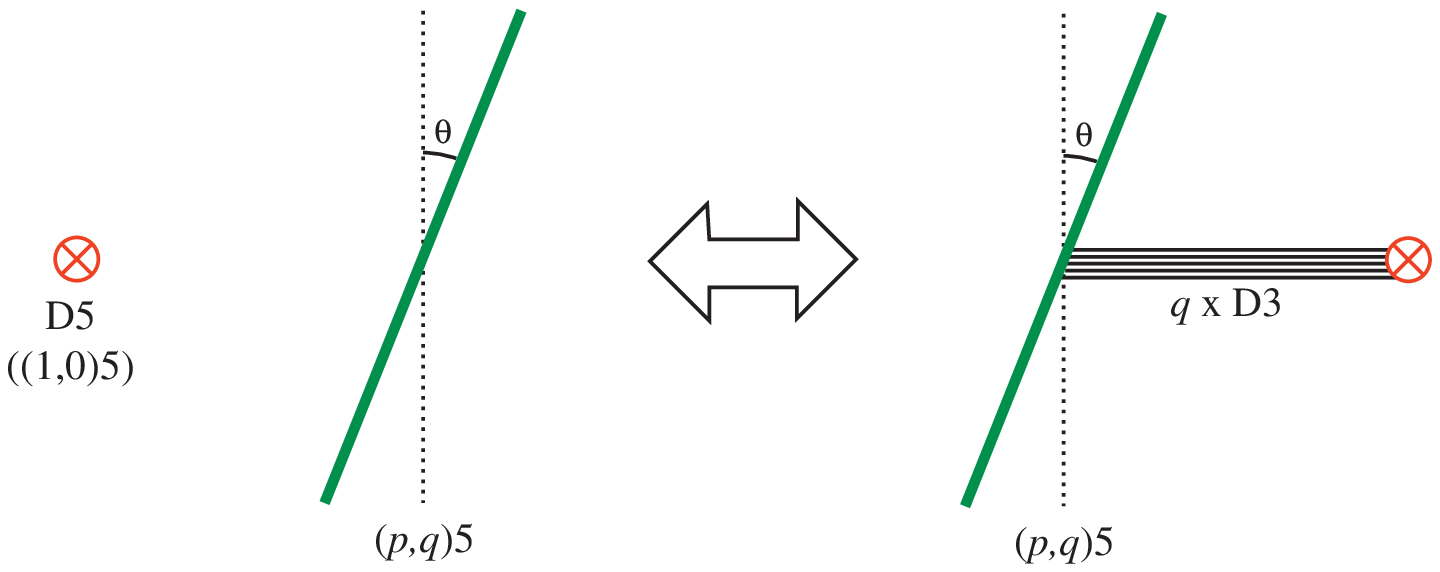}}
\caption{Brane creation with $(p,q)$ fivebrane via $M$-theory.}
\label{creation}
\end{figure}

However, the number of the created M2-branes is different from the
Hanany-Witten case. The number of the M2-branes is determined by how many times
the M5-branes intersect on the wrapped torus. In general, a $(p,q)$5-brane and
a $(p',q')$5-brane intersect $|pq'-qp'|$ times, which is the intersection
number of the cycles of $p\a+q\b$ and $p'\a+q'\b$, on the torus, where $\a$
and $\b$ stand for the two independent cycles of the torus
(see Fig.~\ref{creation}). Therefore, by the exchange of $(p,q)$5-brane and
$(p',q')$5-brane, $|pq'-qp'|$ D3-branes will be created in Type IIB theory.

The above phenomenon is also reversible. Namely, $|pq'-qp'|$ D3-branes between
the $(p,q)$5-brane and $(p',q')$5-brane are annihilated by an exchange of the
5-branes. Moreover, if $N_c$ D3-branes ($0\leq N_c \leq |pq'-qp'|$) are
stretched between the 5-branes, $N_c$ D3-branes are annihilated and
$|pq'-qp'|-N_c$ D3-branes are created after the exchange.

\begin{figure}[htb]
\epsfysize=7cm \centerline{\epsfbox{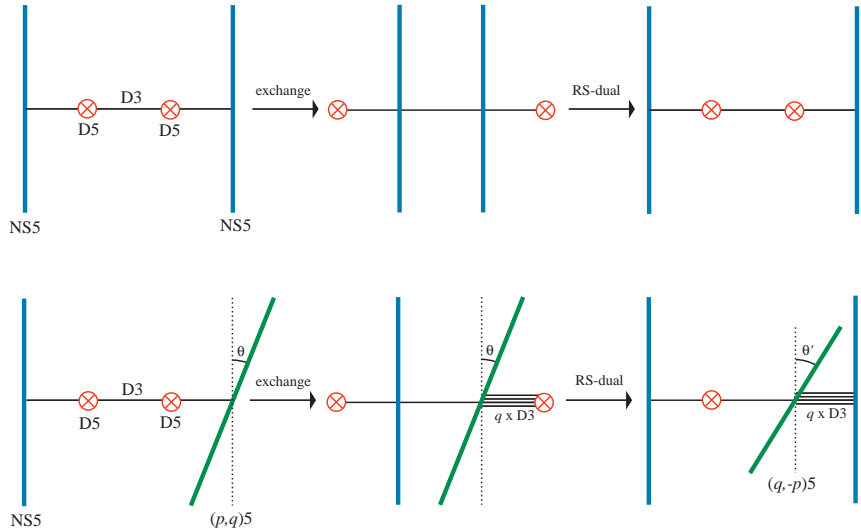}}
\caption{Self-dual theory in the Hanany-Witten configuration and some
particular ``mirror'' in our configuration.}
\label{dual3}
\end{figure}

In the Hanany-Witten configuration~\cite{HW}, there is a simple example of
``mirror'' theory, which is the same theory as the original one. This
self-dual electric theory is a supersymmetric QED with two electrons. After
exchanging positions of two NS5-branes and taking RS-dual in the brane
configuration of this theory, we obtain completely the same configuration
(see the upper row in Fig.~\ref{dual3}).

In our case, we encounter somewhat different situation. Let us now apply the
above particular results of the brane creation to our configuration. We first
move the NS5-brane on the left to the right across one left D5-brane and the
$(p,q)$5-brane on the right to the left across one right D5-brane. Then some
brane creations or annihilations occur and we obtain a configuration that one
D3-brane stretches between two D5-branes and $(q-1)$ D3-branes stretch between
$(p,q)$5-brane and D5-brane (see the middle of the lower row in
Fig.~\ref{dual3}).

We take an RS-dual furthermore. Then we finally end up with a configuration
consisting of a D5-brane and a $(q,-p)$5-brane between two NS5-branes.
At the origin of the Coulomb branch, we can see that one D3-brane stretches
between the left NS5-brane and the middle $(q,-p)$5-brane and $q$ D3-branes
stretch between the middle $(q,-p)$5-brane and the right NS5-brane
(see the right of the lower row in Fig.~\ref{dual3}). After this sequence of
transformations, we find that the resulting ``mirror'' theory is a direct
product of abelian CS theory with a CS coupling $\kappa=q/p$ and $SU(q)$
nonabelian CS theory with $\kappa=q/p$.

The matter contents of the above ``mirror'' theory are not clear and it is
difficult to write down the action for the ``mirror'' of this type explicitly
in the field theoretical point of view. However, the above construction of
``mirror'' theories using branes predicts that there are many nontrivial
``mirror'' relations among various CS theories. We hope that further analysis
of the brane dynamics will shed more light on the understanding
of the duality of these CS theories.

\section{Conclusions and Discussions}

We have investigated in this paper the dynamics of various supersymmetric
gauge theories in 3 dimensions by using brane configurations with
$(p,q)$5-brane in type IIB superstrings. We have given a list of possible
gauge theories, which can also include the CS terms. We have also considered
the vacuum structures of theories with CS terms by using branes.
The dynamics of branes in Type IIB theory tell us that there exist
rich structures of vacua in CS theories. In field theoretical analysis,
it is not easy to understand these vacuum structures of CS theories.
The advantage of the brane probe of field theories is that we can
graphically and easily understand the moduli space of vacua via various
superstring backgrounds. These diagrams of brane configurations also provide
us with easy constructions of ``mirror'' theories.

There are many interesting applications of CS theories themselves. Many
low-dimensional physical systems are described by CS theory as an effective
theory. For example, they include fractional quantum Hall effects or theories
for superconductivity. In these low-dimensional dynamics the dual descriptions
of the theories are also important in order to understand various phenomena in
the systems. Though our investigation is restricted to supersymmetric CS
theories, we expect that our results may apply partially to non-supersymmetric
models as well.

{}From a string theoretical standpoint, the consideration of CS theory as a
worldvolume effective theory on D-branes gives us many informations about
understandings of the Type IIB S-duality. The correspondence between the
S-duality in Type IIB theory and the duality of effective theories on the
branes is mainly studied in a massless sector up to the present. In our brane
configuration with $(p,q)$5-brane massive sectors are crucial since the
effective theory includes CS terms and has a mass gap. We hope that
deeper analyses enable us to understand the nonperturbative dynamics of
string theory.

In our brane configuration, there are some constraints for the masses coming
from the constraints of unbroken supersymmetry. However, when we write down the
action in three-dimensional $N{=}1$ superspace formalism, there is no reason
why arbitrary values of the couplings are not allowed. This means that
there are wider class of possible field theories than those realized on
our brane configurations. This is the case for odd supersymmetry,
that is $N{=}3$ and $N{=}1$ given in our table 2.\footnote{Though it may
appear that there is no constraint on the parameters in $N=3$ configuration,
the CS mass $\kappa_0$ is arbitrary for general $N=3$ field
theories~\cite{ZP}.} The brane realizations of $N=2,4$, on the other hand,
covers all possible parameters. This may be due to the fact
that the fundamental strings appearing in our brane configuration is
orientable, and so only the brane configurations with even supersymmetries
may have the one-to-one correspondence with the supersymmetric field theories.
It would be interesting to find other realizations which exhaust the other
more general $N{=}1,3$ supersymmetric theories.

Another interesting problem is to examine the mirror symmetry for nonabelian
case in more detail, and also the structure of the Higgs branch with matter.
Here the left NS5-brane is originally an M5-brane wrapping along the
$x^2$-direction and the right $(p,q)$5-brane is an M5-brane wrapping on a
torus in M-theory. The M2-brane connects these two types of M5-branes and
the matter M5-branes are along $(1789\natural)$. Since on the torus the
$(p,q)$5-brane intersects $q$ times the $x^{10}$-direction (see Fig.~\ref{f1}),
along which the matter D5-branes exist, our brane constructions suggest that
the M2-branes can only attach to the right $(p,q)$5-brane only at $q$
discrete positions in the $x^{10}$-direction. The investigation of this
problem in the field theory is left for future study.

Finally the class of theories we have examined may contain fractional
strings~\cite{WOS}. If so, it is expected that there may be some connection
between these objects and the fractional statistics which is induced by
the CS terms.

These problems will be further studied elsewhere.

\section*{Acknowledgments}

We would like to thank Boris Zupnik for helpful comments on $N{=}3$
supersymmetric theories. T.K. would also like to thank T. Hotta, I. Ichinose,
M. Ikehara, T. Kuroki and T. Tani for useful discussions.
The work of T.K. is supported  in part by the Japan Society for the
Promotion of Science under the Predoctoral Research Program (No. 10-4361).

%
\newcommand{\NP}[1]{Nucl.\ Phys.\ {\bf #1}}
\newcommand{\AP}[1]{Ann.\ Phys.\ {\bf #1}}
\newcommand{\PL}[1]{Phys.\ Lett.\ {\bf #1}}
\newcommand{\CQG}[1]{Class. Quant. Gravity {\bf #1}}
\newcommand{\NC}[1]{Nuovo Cimento {\bf #1}}
\newcommand{\CMP}[1]{Comm.\ Math.\ Phys.\ {\bf #1}}
\newcommand{\PR}[1]{Phys.\ Rev.\ {\bf #1}}
\newcommand{\PRL}[1]{Phys.\ Rev.\ Lett.\ {\bf #1}}
\newcommand{\PRE}[1]{Phys.\ Rep.\ {\bf #1}}
\newcommand{\PTP}[1]{Prog.\ Theor.\ Phys.\ {\bf #1}}
\newcommand{\PTPS}[1]{Prog.\ Theor.\ Phys.\ Suppl.\ {\bf #1}}
\newcommand{\MPL}[1]{Mod.\ Phys.\ Lett.\ {\bf #1}}
\newcommand{\IJMP}[1]{Int.\ Jour.\ Mod.\ Phys.\ {\bf #1}}
\newcommand{\JP}[1]{Jour.\ Phys.\ {\bf #1}}

\end{document}